\documentclass[a4paper,10pt]{article}
\usepackage[utf8]{inputenc}
\usepackage{amsmath, amsthm, amssymb}
\usepackage{hyperref}
\usepackage{bm}
\usepackage{relsize}
\usepackage{authblk}
\usepackage{dcolumn}
\usepackage{setspace}
\usepackage[normalem]{ulem}
\usepackage{ifpdf}

\title{Dual Scale Invariance Spontaneous Symmetry Breaking and Turbulence}
\author{Itamar Hason\thanks{itamarhason@gmail.com}}
\affil{Raymond and Beverly Sackler School of Physics and Astronomy, \\ Tel Aviv University, Tel Aviv 69978, Israel}
\date{\today}

\begin{document}

\maketitle

\begin{abstract}
We discuss spontaneous symmetry breaking of dual (space and time) scale invariance in the context of turbulence.
\end{abstract}

\section{Introduction}

Turbulence is considered to be the last major unsolved problem of classical physics~\cite{Frisch:1995tu,FalkovichSreenivasan}. It is usually studied in the context of non-relativistic incompressible fluid flows, which are described by the Navier-Stokes (NS) equations
\begin{equation}
\partial_t v_i + v_j \partial_j v_i = -\partial_i P + \nu \partial_{jj} v_i + f_i
\qquad
\partial_i v_i = 0,
\end{equation}
where $v_i$ is the fluid velocity, $P$ is the pressure (divided by the constant density), $\nu$ is the fluid viscosity and $f_i$ some external force. Due to the strong non-linearity of the NS equations, the study of turbulent flows has focused on the statistical properties of the solutions. There are strong indications that these properties exhibit some degree of universality but a full analytical understanding is still missing.

Solutions to the NS equations are characterized by the Reynolds number, defined by:
\begin{equation}
Re = \frac{L v}{\nu},
\end{equation}
where $L$ is the characteristic length scale of velocity difference and $v$ is the typical velocity difference. Flows with $ Re \ll 1$ are dominated by the linear terms and are laminar while flows with $Re \gg 100$ are dominated by the non-linear term and become turbulent. One defined also the viscous scale $l\sim \left(\nu^3/\epsilon\right)^{1/4}$.

A statistical turbulent steady-state can be maintained by some stationary random external force constantly being applied at the scale $L$. One usually studies the longitudinal structure functions of order $n$:
\begin{equation}\label{tur:structurefuncs}
S_n (r) \equiv \left\langle \left( \delta \vec{v}(\vec{r}) \cdot \frac{\vec{r}}{r} \right)^n \right\rangle .
\end{equation}
where $ \delta \vec{v} (\vec{r})$ is the velocity difference on displacement $\vec{r}$. A very well known, and one of few exact results for the statistics of turbulence, is Kolmogorov's four-fifths law \cite{Kol:1941a}
\begin{equation}\label{tur:Kol45law}
S_3 (r) = - \frac{12}{d\left(d+2\right)} \epsilon r,
\end{equation}  
where $d$ is the number of space dimensions, $\epsilon$ is the mean rate of energy dissipation per unit volume and $l\ll r\ll L$ ($Re \rightarrow \infty$).
 
Kolmogorov \cite{Kol:1941b} assumed that in this limit the velocity statistics in the inertial range are scale invariant. This assumption and dimensional analysis then lead to the conclusion that:
\begin{equation} \label{eq:univeral_scaling}
S_n (r) \propto (\epsilon r )^{n/3},
\end{equation}
which is known as \emph{K41 scaling}. Experimental and numerical evidence suggests that turbulent flows in the inertial range do exhibit universal behaviour, such that the structure functions scale as power laws:
\begin{equation}\label{eq:anomscaling}
S_n(r) \propto r^{\xi_n} \ .
\end{equation}
However, in $d\ge3$ space dimensions, $\xi_n$ is not linear in $n$, and deviates from the prediction of $\xi_n = n/3 $ obtained from the self-similarity (or scale invariance) assumption (except for $n=3$, for which \eqref{tur:Kol45law} holds). The calculation of the anomalous exponents $\xi_n$ remains an open problem.

The Euler equation, which is the Navier Stokes equation with zero viscosity, is a \emph{dual scale} invariant theory -- it's scale invariant independently in both the space dimensions and the time dimension. The NS equations are of-course not dual scale invariant, and in the limit of zero viscosity the theory is still anomalous and the scale symmetry is broken to $z=2/3$, as can be seen from \eqref{tur:Kol45law} (which should be valid at $\nu\rightarrow0$ but not  vanishing). This breaking can be thought to be spontaneous \cite{Oz:2017ihc} and in that sense, the study of turbulence has strong connections with the study of spontaneous symmetry breaking and anomalies in non relativistic field theories (see \cite{Arav:2017plg} for a related discussion of non-relativistic scale symmetry breaking and anomalies).

Eling and Oz proposed \cite{Eling:2015mxa} an analytical formula for the anomalous scaling exponents of inertial range structure functions in incompressible fluid turbulence which is a Knizhnik-Polyakov-Zamolodchikov (KPZ)-type relation \cite{Knizhnik:1988ak,Rhodes:2013iua} and is claimed to be valid in any number of space dimensions. In this formula the anomalous exponents are the K41 exponents \eqref{eq:univeral_scaling} dressed via a coupling to a lognormal random geometry:
 \begin{equation} \label{eq:kpz_anomalous_exponents}
  \xi_n - \frac{n}{3} = \gamma^2\left(d\right)\xi_n\left(1-\xi_n\right) \ .
 \end{equation}
This formula has one real parameter $\gamma$ that depends on the number of space dimensions. The use of the KPZ formula is inspired by the fluid/gravity correspondence \cite{Bhattacharyya:2008kq, Eling:2009pb, Eling:2010vr}, the fluid is thought to be coupled to random geometry. A derivation of \eqref{eq:kpz_anomalous_exponents} was suggested in \cite{Oz:2017ihc}, relying on the effective action for the spontaneous symmetry breaking pattern we will discuss in this paper.

As mentioned above, the turbulence problem exhibits a two-dimensional scale symmetry which is thought to be spontaneously broken to a regular one dimensional Lifshitz scaling. The above proposal and this symmetry breaking mechanism suggest we should study the spontaneous symmetry breaking of this dual scale, or ``arbitrary $z$'' scale symmetry and the Nambu-Goldstone mode that should exist in such theory. A particle with a logarithmic correlation function may serve as the mediator for the random geometry process needed to establish the KPZ relation, therefore, a logarithmic-correlated NG mode will be of a great interest.

\section{Dual Scale Invariant Effective Action}
We are looking for an effective action which would be scale invariant under an \emph{arbitrary} scale symmetry, i.e., the whole set of transformations
\begin{equation}
 \left({ \vec{x},t} \right) \rightarrow \left( { \lambda \vec{x}, \lambda^z t } \right)
\end{equation}
for arbitrary $z$.
We can also parametrize these tranformations differently by
\begin{equation}
 \left({ \vec{x},t} \right) \rightarrow \left( { \lambda_1 \vec{x}, \lambda_2 t } \right) \ .
\end{equation}
We are interested in the dilaton \emph{steady-state effective action} which we expect to be a space(-only) integration of a lagrange density $L$.
A natural guess would be
\begin{equation} \label{random_geometry_lagrangian}
 L = \tau \left(\nabla^2\right)^{d/2}\tau
\end{equation}
which induces an effective action invariant under a general tranformation $\vec{x} \rightarrow \lambda \vec{x} $, regardless of $z$ (or $\lambda_2$). It would be nice if this was the only option since a particle with such action in $d$ space dimensions has a logarithmic correlation function, the problem is that by declaring the tranformation rule $\tau \rightarrow \tau + \sigma$ for $\lambda = e^\sigma$, we get that we can also have
\begin{equation}
 L = e^{-\alpha\tau} \left(\nabla^2\right)^{\beta/2}\tau
\end{equation}
as long as $\alpha+\beta=d$
 
Note that the original symmetry group had 2 degrees of freedom -- $\lambda$ and $z$, or $\lambda_1 $ and $\lambda_2$, but the symmetry we required the dilaton lagrangian to obey was parametrized by only one of them $\lambda$ (or $\lambda_1$). This seems like a simplification of the requirement and we will see that although we don't introduce time dependence or time integration (by the steady-state assumption), there is a way to require both $\lambda$ and $z$ (or $\lambda_1$ and $\lambda_2$), and therefore getting stricter requirements.
 
Note also that usually when one deals with Lifshitz scaling with fixed dynamical exponent $z$, the dilaton transformation is defined to be
\begin{equation}
 \tau\left(\vec{x},t\right) \rightarrow \tau \left( {e^{-\sigma} \vec{x} , e^{-z\sigma}t } \right) + \sigma 
\end{equation}
such that the dilaton is able to directly compensate for space variations, by the $\sigma$ contribution added to $\tau$ which corresponds to the space tranformation $\vec{x} \rightarrow e^{-\sigma}\vec{x}$. However, we can suggest a different representation:
\begin{equation}
 \tilde\tau\left(\vec{x},t\right) \rightarrow \tilde\tau \left( {e^{-\sigma/z} \vec{x} , e^{-\sigma}t } \right) + \sigma 
\end{equation}
such that the tilded dilaton $\tilde\tau$ is able to directly compensate for time variations since in this representation the added $\sigma$ contribution corresponds to the time transformation $t \rightarrow e^{-\sigma}t$.

By moving to this representation, it is justified to require the space part of the dilaton lagrangian to be invariant under the two parameters tranformation:
\begin{equation} \label{new_dilaton_transformation}
 \tilde\tau \rightarrow \tilde\tau + \sigma,  \vec{x} \rightarrow \lambda\vec{x}
\end{equation}
with no connection between $\sigma$ and $\lambda$, which are in fact related by the hidden (and arbitrary) $z$. In this representation, $\sigma$ defines the time scaling, and $\lambda$ defines the space scaling, so that it is clear that the lagrangian should be invariant under the dual scaling -- space and time independently.
 
Suppose we are left with an unbroken scale symmetry with some specific $z$, so that $\sigma_{1,2} = z_{1,2}\sigma$, i.e. $z = z_2/z_1$ is unbrokem while other $z$'s are broken. We should expect $\tau$ to be invariant under the unbroken symmetry, thus, up to overall normalization factor,
\begin{equation}
 \tau \rightarrow \tau + \alpha\sigma_1 + \beta\sigma_2
\end{equation}
with 
\begin{equation}
 \alpha z_1 + \beta z_2 = 0 \ .
\end{equation}
In our case $z=2/3$ so we have $\tau \rightarrow \tau - 2\sigma_1 + 3\sigma_2$.

In general, for when we have $\tau \rightarrow \tau + \alpha\sigma_1 + \beta\sigma_2$ with nonzero $\beta$, the possible leading order space only lagrangian is only $\tau \left( \nabla^2 \right) ^ {d/2}\tau$ because any $\tau$ exponent would allow an arbitrary time scaling to ruin the symmetry since the exponent term would scale while all space derivatives wouldn't.

We can have higher order terms. Every $\tau$ should be accompanied by a space derivative, and the total dimension (number of derivatives) should be $d$. So for example $(\partial\tau)^d$ is a possibility. Rotation invariance requires of course all derivative indices to be contracted in pairs, which is a constraint. This constraint seems to forbid all local terms in the odd space dimensionality case, we may keep $\tau \left( \nabla^2 \right) ^ {d/2}\tau$ but recognize the fact that it is non-local. For even dimensionality we have more terms the higher the dimensionality is. For example, in 2d we have only $\tau \Delta \tau$ (which is equivalent to $\partial\tau \partial\tau$), in 4d we have $\tau \Delta^2 \tau$, $(\partial\tau \partial\tau) \Delta \tau$, $(\partial\tau \partial\tau)^2$.

\section*{Acknowledgments}
We would like to thank Igal Arav and Yaron Oz for valuable discussions, and specifically Yaron Oz for introducing this problem to me. This work is supported in part by the I-CORE program of Planning and Budgeting Committee (grant number 1937/12), the US-Israel Binational Science Foundation, GIF and the ISF Center of Excellence.

\end{document}